\documentclass[preprint2]{aastex}

\slugcomment{submitted to ApJ 2002 April 26; revised 2002 June 14; revised 2002July 29}

\shorttitle{Microlensing and Suppressed Saddlepoints}
\shortauthors{Schechter and Wambsganss}

\begin{document}


%
%
%
\title{Quasar Microlensing at High Magnification and the Role
of Dark Matter: Enhanced Fluctuations and Suppressed Saddlepoints}


\author{Paul L. Schechter}

\affil{Department of Physics, Massachusetts Institute 
of Technology, 77 Massachusetts Avenue, Cambridge, MA 02139}
\affil{Institute for Advanced Study, Einstein Drive, Princeton, NJ 08540-0631}

\and

\author{Joachim Wambsganss}

\affil{Universit\"at Potsdam, Institut f\"ur Physik, Am Neuen Palais 10,
14467 Potsdam, Germany }




\begin{abstract}
Contrary to naive expectation, diluting the stellar component of the
lensing galaxy in a highly magnified system with smoothly distributed
``dark'' matter {\it increases} rather than decreases the microlensing
fluctuations caused by the remaining stars.  For a bright pair of
images straddling a critical curve, the saddlepoint (of the arrival
time surface) is much more strongly affected than the associated
minimum.  With a mass ratio of smooth matter to
microlensing matter of 4:1, a saddlepoint with a macro-magnification
of $\mu = 9.5$ will spend half of its time more than a magnitude
fainter than predicted.  The anomalous flux ratio observed for the
close pair of images in MG0414+0534 is a factor of five more likely
than computed by Witt, Mao and Schechter if the smooth matter
fraction is as high as 93\%.  The magnification probability histograms
for macro-images exhibit distinctly different structure that varies
with the smooth matter content, providing a handle on the smooth matter
fraction.  Enhanced fluctuations can manifest themselves either in the
temporal variations of a lightcurve or as flux ratio anomalies in a
single epoch snapshot of a multiply imaged system.  While the
millilensing simulations of Metcalf and Madau also give larger
anomalies for saddlepoints than for minima, the effect appears to be
less dramatic for extended subhalos than for point masses.  Morever,
microlensing is distinguishable from millilensing because it will
produce noticeable changes in the magnification on a time scale of a
decade or less.
\end{abstract}




\section{INTRODUCTION}

Simple gravitational lens models have proven remarkably successful at
reproducing the observed positions of multiply imaged objects.  But
the agreement between the flux ratios predicted by these models and the
flux ratios actually observed is mediocre at best.  At the level of
several tenths of a magnitude, far worse than the observational
accuracy, the models begin to fail.  

These shortcomings have variously been attributed to the effects of
intervening dust (Lawrence et al. 1995), microlensing by the stars
which comprise the lens (Witt, Mao and Schechter 1995, hereafter WMS)
or millilensing by galactic substructure (Mao and Schneider 1998;
Metcalf and Madau 2001; Chiba 2002; Dalal and Kochanek 2002).

The discrepancies are especially obvious in a number of quadruply
imaged systems that have a close pair of highly magnified images
straddling a critical curve.  On fairly general grounds (e.g. Gaudi
and Petters 2002), these pairs of images are expected to be roughly
equal in brightness, with model dependent uncertainties of order 10\%
(Metcalf and Zhao 2001).  The best known discrepancy is that of
MG0414+0534 (Hewitt et al. 1992; Schechter and Moore 1993) for which
the observed optical $A_2/A_1$ flux ratio is a factor of two smaller
than the 1:1 ratio that is both theoretically expected and observed at
radio wavelengths (Trotter et al. 2000).  Taking the lensing potential
to be isothermal, and assuming that the lensing mass was comprised
entirely of stars, it was argued by    Witt, Mao and Schechter that
microlensing might, on rare occasions, produce so dramatic a flux
ratio discrepancy.  They also noted that the expected fluctuations are
larger for images that are saddlepoints (as opposed to minima) of the
arrival time surface.

Our interest in anomalous flux ratios was reawakened by four recent
discoveries (Reimers et al. 2002; Inada et al., in preparation; Burles
et al., in preparation; Schechter et al., in preparation) of quadruple
systems with similarly anomalous pairs of images.  In three cases (and
possibly in the fourth), the fainter image is a saddlepoint and the
brighter image is again a minimum, as was the case in MG0414+0534.
While there has been, as yet, no systematic survey, the present paper
proceeds on the working hypothesis that this phenomenon may be fairly
general.

As in WMS, we start with the assumption that the lensing potential is
isothermal, which requires a variable mass to light ratio if not
``dark'' matter.  We find that the expected brightness fluctuations for
highly magnified macro-images of quasars are {\it enhanced} as one
relaxes the assumption that the lensing galaxy is comprised entirely
of stars and admits a substantial, smoothly distributed ``dark''
component.\footnote 
{For economy of hypothesis we take our smoothly distributed component
to be identical with the ``dark'' matter, the presence of which is
inferred from multiple lines of reasoning.  This would be on firmer
ground if more than 50\% of the matter along the line of sight can be
shown to be both ``dark'' and smoothly distributed.}  While the
fluctuations for minima are only somewhat larger, those for
saddlepoints are very much larger.  In both cases the magnification
distributions become asymmetric, with a substantial probability that
saddlepoints will be very much fainter than predicted by macro-models.

The idea of using microlensing to set limits on the smoothly
distributed matter content of galaxies has been explored before by
Webster et al. (1991) and Lewis and Irwin (1996).  The present
argument builds on findings by Deguchi and Watson (1987, 1988) and
Seitz, Wambsganss and Schneider (1994) that as one increases the
surface density of microlenses -- at fixed shear -- the microlensing
fluctuations at first increase and then decrease as one approaches
infinite magnification.  The same holds true for increasing the
dimensionless surface density and setting the shear approximately
equal to it (WMS).

Here we focus on the fluctuations resulting from a different slice
through the plane spanned by surface density and shear.  This locus
traces the values of the ``effective'' surface density $\kappa^{\rm
eff}$ and ``effective'' shear $\gamma^{\rm eff}$ (Paczy{\'n}ski 1986)
obtained as one gradually substitutes a smoothly distributed surface
mass density $\kappa_{\rm c}$ for the clumpy, stellar mass density
$\kappa_*$.

With the benefit of hindsight, both effects described below     can be
reconstructed from previously published results.  Our finding that
adding smoothly distributed matter enhances fluctuations follows
immediately from Pacy{\'n}ski's (1986) scaling and the demonstration
that fluctuations go to zero at infinite magnification.  Our finding
that saddlepoints and minima behave very differently has antecedents
in the microlensing simulations of Wambsganss (1992), Witt, Mao and
Schechter (1995) and Lewis and Irwin (1995, 1996), and leads
ultimately back to the work of Chang and Refsdal (1979).

In \S2 we review some basics of macro- and microlensing.  In \S3 we
consider a toy model for the two effects we seek to explain.  In \S4
we examine microlensing simulations for parameters appropriate to
quadruple systems.  In \S5 we compare results of these simulations to
the specific case of MG0414+0534.  In \S6 we discuss how ensembles of
systems might be used to estimate the smoothly distributed matter
fraction in galaxies.  In \S7 we consider the consequences of our
findings for millilensing by sub-halos and fitting lens models, and we
examine the effects of relaxing our assumption of isothermality.

\section{MACRO- AND MICROLENSING BASICS}

\subsection{Macro-models}

Following Witt et al. (1995), we restrict ourselves to modelling macro-lenses
as singular isothermal spheres (SIS) with an external tidal field.
Images form at minima and saddlepoints of the Fermat travel-time
surface (Blandford and Narayan 1986).  Formally there is also an image
at or near the central maximum, but for the SIS model this maximum is
infinitely demagnified.  An approximate relation is derived in WMS
between the dimensionless surface density (the convergence),
$\kappa_{\rm tot}$, and the combined effect of all tides -- the shear,
$\gamma$ -- at the positions at which images form in such a model,
\begin{equation}
\gamma \approx 3\kappa_{\rm tot} - 1 \quad .
\end{equation}
The magnification $\mu_{\rm macro}$ of a macro-image is given by the inverse
of the product of the eigenvectors of the curvature matrix,
\begin{equation}
\mu_{\rm macro} = {1 \over [(1-\kappa_{\rm tot}) + \gamma] [(1-\kappa_{\rm tot})  - \gamma]}
\quad .
\label{eq-mag}
\end{equation}
An image is a minimum of the travel-time surface if $1 - \kappa_{\rm tot}
- \gamma > 0$, a saddlepoint if $1 - \kappa_{\rm tot} - \gamma < 0$, and a
maximum if $1 - \kappa_{\rm tot} + \gamma < 0$.  While maxima and saddlepoints
may have $|\mu_{\rm macro}| < 1$, indicating demagnification, minima must
always be magnified.

For parameters typical of quadruple systems, the brighter images are
magnified by factors $\mu_{\rm macro} \approx 5-20$.  The typical
close pair of macro-images in a quad might have $(\kappa_{\rm
tot},\gamma) = (0.475, 0.425)$ for the minimum and $(0.525, 0.575)$
for the saddlepoint, giving them magnifications of $\mu_{\rm minimum}
= 10.5$ and $\mu_{\rm saddle} = -9.5$, respectively, with the negative
magnification indicating the parity flip associated with a
saddlepoint.  We shall use these values for simulations presented in
\S4.

\subsection{Micro-images}

Introducing small scale perturbations in the lens potential produces
new hills, valleys and ridges that are stationary points of the
Fermat surface, thereby producing micro-images.  For a point source, what we
think of as a coherent macro-image will then be comprised of many such
micro-images.  Paczy{\'n}ski's (1986) Figure 1 illustrates this especially
well.  For point mass perturbers the micro-images are either minima or
saddlepoints.  

For both macro-saddles and macro-minima, there is a nearly infinite
number of faint micro-saddlepoints, one for every star.  In general
there are extra negative/positive parity pairs of micro-images, with
the mean number increasing with increasing macro-magnification.
Wambsganss, Witt and Schneider (1992) give an expression for the
dependence of the mean number of extra image pairs on the density of
microlenses in the absence of external shear.  Depending upon the
accidental distribution of microlenses, a macro-saddlepoint may or may
not have micro-minima.  By contrast, a macro-minimum must have at
least one.  As with macro-images, micro-saddlepoints may be highly
demagnified but micro-minima must have magnifications greater than
unity.

\section{SUPPRESSION OF SADDLEPOINTS: EXPLANATIONS}

Witt et al. (1995) noticed that the fluctuations in their simulations
of microlensing were larger for saddlepoints (standard deviation
$\sigma_{\rm saddle} \sim 0.9$ mag) than for minima ($\sigma_{\rm
minimum} \sim 0.6$ mag).  They offered the following explanation:
\begin{quotation}
The macro-images of positive parity have smaller fluctuation because
their magnifications must be larger than unity whereas the macro-images
of negative parity have no lower limit in magnification.
\end{quotation}
This sounds plausible but for the simulations presented in the next
section we find that the fluctuations are much larger for the
saddlepoints even though the magnifications rarely, if ever, dip below
unity.  Here we offer a toy model which we believe captures the
elements of the effect.  Readers who prefer not to be toyed with may
wish to skip to \S4 and then return to it.

\subsection{A Toy Model}

We consider the extreme case of a lens in which all but an
infinitesimal fraction of the mass is in a smoothly distributed
component.  We then introduce a single point mass perturber and
examine its effects on the total magnification of macro-saddlepoints
and macro-minima.  Our treatment is essentially that of Chang and
Refsdal (1979; 1984), though with different emphasis and notation.

A macro-image is characterized by a local value of the convergence
$\kappa_{\rm tot}$ and the shear, $\gamma$.  In the absence of the
perturber, the magnification is given by equation (\ref{eq-mag})
above.  For the highly magnified images typical of quadruple systems,
the curvature matrix for a saddlepoint (or minimum) has a deep
minimum directed approximately radially outward from the lens center
and a broad maximum (or minimum) in the tangential direction.  In the
absence of microlenses each macro-image is comprised of exactly one
micro-image with the exact properties expected for the macro-image.

For both cases, macro-minimum and macro-saddlepoint, we introduce a
point mass perturber which, for simplicity, we place directly along
the line of sight to the macro-image.  The macro-minimum is split into
4 micro-images: two micro-minima along the tangential direction 
and two micro-saddlepoints along the radial direction,
with magnifications
\begin{equation}
\mu_{\rm micro} = \left\{ \begin{array}{ll}
{1 \over 4\gamma[(1 - \kappa_{\rm tot}) - \gamma]} & {\rm (minimum)}
\\[2pt]
{1 \over 4\gamma[(1 - \kappa_{\rm tot}) + \gamma]} & {\rm (saddle)}
\end{array} \right. \quad .
\end{equation}
The same perturber along the line of sight to the macro-saddlepoint
produces only the two micro-saddlepoints along the radial direction, with
identical magnifications to those in the case of the macro-minimum.
Summing over the micro-images we have 
\begin{equation}
\mu_{\rm toy} = \left\{
\begin{array}{ll}
{ (1 - \kappa_{\rm tot}) \over
\gamma} \mu_{\rm macro}
& {\rm (minimum)} 
\\[5pt]
{[(1 - \kappa_{\rm tot}) - \gamma]  
\over
2 \gamma}  \mu_{\rm macro}
& {\rm (saddle)}
\end{array}
\right.  \quad .
\end{equation}
Taking $\gamma \approx \kappa_{\rm tot}$, typical of the SIS,
and we have 
\begin{equation}
\mu_{\rm toy}^{\rm SIS} \approx
\left\{
\begin{array}{ll}
{ (1 - \kappa_{\rm tot}) \over
\kappa_{\rm tot}} \mu_{\rm macro}
& {\rm (minimum)} 
\\[5pt]
{1 \over 2\kappa_{\rm tot}}
& {\rm (saddle)}
\end{array}
\right.  \quad .
\end{equation}
For highly magnified SIS images, $\kappa_{\rm tot} \approx 1/2$.  The
perturbed and unperturbed macro-minima then have nearly the same
magnification.  By contrast the perturbation undoes the
macro-magnification of the saddlepoint, returning it to roughly its
brightness without any macro-magnification.  This is the essence of
our explanation for the different behavior of macro-saddlepoints and
macro-minima.

\subsection{Variations}

The probability of a such direct hit is vanishingly small.
Moving the perturber away from the macro-minimum, the total
magnification increases, slowly at first but then rapidly.  A pair of
micro-images merge and the magnification drops to something close to the
predicted macro-magnification.  As the perturber moves off to
infinity, a single micro-minimum is left that asympotically
approaches the predicted macro-magnification.

Moving the perturber away from the macro-saddlepoint along the radial
direction, one saddlepoint decreases in brightness while the other
increases, asymptotically approaching the macro-magnification.  Moving
it away along the tangential direction, the total magnification
increases at first slowly but then dramatically with the creation of a
new pair of micro-images.  It then falls dramatically after the newly
created micro-minimum merges with one of the original
micro-saddlepoints.  One of the two remaining micro-saddlepoints
asymptotically approaches the macro-magnification.

In both cases, moving the perturber off the line of sight introduces
some additional fluctuations in the sum of the micro-images, but it
does not change the fundamental behavior.  Most of the time the
perturbed macro-minimum is {\it slightly} brighter than it otherwise
would have been, while for much of the time the macro-saddlepoint is
{\it very much} fainter than it would have been.

What happens as we increase the mass fraction in microlenses?  As long
as the microlenses are sparsely distributed, they don't interfere with
each other.  The average number of extra micro-minima is less than
unity and their magnification distribution looks much as it did for
the single perturber.  But as the microlens density increases, higher
order caustics begin to form, the average number of extra micro-minima
grows, and their magnifications are influenced by local density
fluctuations rather than the global parameters.  The total
magnification is proportional to the number of micro-minima
(Wambsganss et al. 1992), with fractional fluctuations decreasing as
$1/\sqrt{N}$.  The fractional fluctuations therefore have a maximum
somewhere between 100\% smoothly distributed matter and 100\% clumpy
(microlensing) matter.

Though macro-maxima are only rarely seen in lensed systems (and not in
the systems considered here) our toy model helps in their
interpretation as well.  A perturber directly along the line of sight
to a macro-maximum makes the curvature of the maximum infinite,
reducing the magnification to zero.  Moving the perturber away from
the macro-maxium at first produces no change -- no flux.  At some
point a new micro-maximum/micro-saddlepoint pair is created.  The
micro-maximum moves toward the     macro-maximum and the
micro-saddlepoint moves closer to the perturber.

\section{MICROLENSING SIMULATIONS}

The handwaving of the preceding section can be tested by numerical
simulations of microlensing, as pioneered by Paczy\'nski (1986),
Schneider \& Weiss (1987), Kayser, Refsdal, Stabell (1987), and
elaborated by Wambsganss (1990), Witt(1993) and Lewis et al. (1993).

\subsection{Choice of Parameters}

At first it might seem that the space of simulations should be three
dimensional, as parameterized by the shear, $\gamma$, the convergence
provided by smoothly distributed matter, $\kappa_{\rm c}$, and the
convergence contributed by a clumpy stellar component, $\kappa_*$,
with $\kappa_{\rm c} + \kappa_* = \kappa_{\rm tot}$. 
Paczy{\'n}ski (1986) has shown
that there is a scaling such that for any choice of these 
there is an equivalent model with an effective shear
\begin{equation}
\gamma^{\rm eff} = {\gamma \over (1-\kappa_{\rm c})}
\end{equation} 
and an effective convergence, 
\begin{equation} 
\kappa^{\rm eff}_* = {\kappa_* \over (1-\kappa_{\rm c})} \quad .
\end{equation}
but no smooth component.  One must also scale each dimension of the
source plane by a factor $(1-\kappa_{\rm c})^{-1}$ and the resulting
magnifications by a factor $(1-\kappa_{\rm c})^{-2}$.  To investigate
the effect of substituting continuous matter for stellar matter, one
starts with a microlensing simulation in which $\kappa_* = \kappa_{\rm
tot}$ and $\gamma$ are given by the macro-model, then gradually
increases $\kappa_{\rm c}$ and decreases $\kappa_*$ keeping
$\kappa_{\rm tot}$ constant.

In Figure 1 we show the $\kappa^{\rm eff}_* - \gamma^{\rm eff}$ plane.
Macro-minima lie in the shaded region in the lower left; macro-maxima
lie in the cross-hatched region to the lower right; macro-saddlepoints
lie in the remaining triangular region.  The dotted line shows the
$\gamma^{\rm eff} = \kappa^{\rm eff}$ locus, which is appropriate to
the unperturbed SIS model in the absence of smoothly distributed
matter.  The hyperbolae mark lines of constant $\mu^{\rm eff}$,
defined by
\begin{equation}
\mu^{\rm eff} = {1 \over [(1-\kappa^{\rm eff}_*) + \gamma^{\rm eff}] 
                       [(1-\kappa^{\rm eff}_*) - \gamma^{\rm eff}]}
\quad .
\end{equation}
Note that $\mu^{\rm eff} = (1 - \kappa_{\rm c})^2 \mu_{\rm tot}$ and
is equal to the total magnification only in the absence of smoothly
distributed matter.  The symbols show the effect of increasing the
smoothly distributed matter fraction in a pair of models.  They
terminate on the $\kappa_*^{\rm eff} = 0$ axis for 100\% smooth
matter.

We have carried out two pairs of microlensing simulations.  The first
of these follows a minimum and a saddlepoint with total magnifications
$\sim 10$ as we add increasing proportions of smoothly distributed
matter, with $\kappa^{\rm eff}$ and $\gamma^{\rm eff}$ given by the
symbols in Figure 1.  The second was chosen to permit direct
comparison with the WMS models for MG0414+0534, which have total
magnifications $\sim 25$, and are also carried out with increasing
proportions of smoothly distributed matter.  The relevant parameters
are given in Table 1.

\begin{deluxetable}{rccr} 
\tablecolumns{4} 
\tablewidth{0pc} 
\tablecaption{Simulation Parameters} 
\tablehead{ 
\colhead{ID}   &  \colhead{$\kappa_{\rm tot}$} & \colhead{$\gamma$}
& \colhead{$\mu_{\rm tot}$}
}
\startdata
 M10 & 0.475 & 0.425 &  10.5 \\
 S10 & 0.525 & 0.575 &  -9.5 \\
 M25 & 0.472 & 0.488 &  24.2 \\
 S25 & 0.485 & 0.550 & -26.8 \\
\enddata 
\end{deluxetable} 

\subsection{Image Magnification $|\mu_{\rm tot}| \sim 10 $}

The character of the brightness fluctuations is revealed by mapping
the magnification of the source as a function of its position in the
source plane (Wambsganss, Paczy\'nski \& Schneider  1990).  
In Figure \ref{fig-color} we show the
magnification maps for the ``typical'' macro-minimum (on the left) and
macro-saddlepoint (on the right) with smooth matter fractions of 0\%,
$\sim 85$\% and $\sim 98$\%, from top to bottom.

Figure \ref{fig-histo} presents magnification histograms computed for
each of these.  These are the top, middle and bottom panels.  In the
second and fourth panels we include intermediate histograms for smooth
matter fractions $\sim 75$\% and $\sim 95$\%.  For the sake of
comparison with our toy model, we discuss the 98\% smooth matter model
first and proceed to higher densities of microlenses.

Both the magnification maps and the histograms for the 98\% smooth
matter cases confirm the qualitative results of our toy model.  For
the macro-minimum, the magnification is mostly just that given by the
smooth macro-model.  There are isolated diamond shaped caustics that
surround ``plateaus'' of slightly higher magnification which
correspond to the 4 micro-image regions described in the toy model.
They have one extra micro-minimum.  The sharp transitions between the
typical magnification and the plateaus are the caustics, marking the
points at which a new micro-saddlepoint/micro-minimum pair is created
or annihilated.  Students of stellar and planetary microlensing will
recognize these configurations as the magnification map of a planet
lying just outside the Einstein ring of its parent star (Chang and
Refsdal 1979, 1984; Mao \& Paczy\'nski 1991; Wambsganss 1997).

For the macro-saddlepoint, the magnification is again mostly just that
given by the smooth macro-model.  But there are occasional ``lagoons''
of very low magnification, bracketted by small triangular ``calderas''
of high magnification.  The lagoons appear whenever a microlens lies
directly along the line of sight.  The calderas mark the 4 micro-image
regions described in the toy model, and have one extra micro-minimum.
The lagoons have no micro-minima.  This lagoon/twin-triangle
configuration corresponds to the magnification map of a planet inside
the Einstein ring of its parent star (Chang and Refsdal 1979, 1984;
Wambsganss 1997).

For the 98\% (and 95\%) cases the histograms for macro-minima and
macro-saddlepoints are quite different.  Though both are narrow, the
saddlepoint histogram is broader and shifted toward high magnifications
rather than low magnifications.  Moreover it has a very long tail,
over the length of which it is quite flat, extending faintward toward
unit magnification.  It will therefore not be surprising if, with more
microlenses, the magnification histogram for the macro-saddlepoint is
broader than that for the macro-minimum.

The magnification maps for the 85\% cases bear some resemblance to the
98\% smooth matter maps, but are far more complex.  For the
macro-minimum the diamond caustics now begin to overlap, sometimes
multiple times, producing successively higher plateaus.  The area
completely outside the caustics, regions that produce only one
micro-minimum, now have clearly lower than average magnification.  The
corresponding magnification histogram shows two distinct peaks
associated with regions of one and two micro-minima (cf. Rauch et
al. 1992).  In the 75\% smooth matter histogram this bifurcation is even
more pronounced.

For the macro-saddlepoint the lagoons, regions with no micro-minima,
have now grown to the point at which they dominate the magnification
map.  The triangular calderas have also grown, to the point at which
their caustics sometimes cross, producing regions with two, three and
four micro-minima.  The magnification histogram shows two
distinct peaks, corresponding to regions with zero and one
micro-minima, respectively, and perhaps a plateau corresponding to
regions with two micro-minima.  The histogram has a long tail toward
high magnifications and drops abruptly very near the minimum
magnification allowed by our toy model, $\mu_{\rm tot}=1$.  Again, the
75\% smooth matter histogram shows a yet more pronounced bifurcation.

In the 0\% smooth simulations the macro-minimum and macro-saddlepoint
have begun to resemble each other.  For the macro-saddlepoint, the low
magnification lagoons, still with no micro-minima, have been crowded
out by the expanding web of caustics.  Most of the magnification map
is covered by regions with large numbers of micro-minima.  The web of
caustics has also expanded for the macro-minimum.  The region outside
all caustics looks much the same as for the macro-saddlepoint, though
these regions still produce one micro-minimum.

We get the general impression that the diamond and
lagoon/twin-triangle configurations are the building blocks from which
the higher density configurations are constructed.  We start out with
relatively isolated features.  As the surface density of perturbers
increases, the fluctuations increase as the building blocks begin to
cover the source plane.  Increasing the density of perturbers yet
further, the building blocks overlap and begin to average out,
decreasing the fractional amplitude of the fluctuations.

\subsection{Image Magnification $|\mu_{tot}| \sim 25 $}

Witt et al. (WMS) considered the specific case of MG0414+0534, a
quadruple system for which the     flux ratio of the close bright pair
of images, $A_2/A_1$, was 0.9 in the radio (Katz and Hewitt 1993) and
$0.45 \pm 0.06$ in the optical (Schechter and Moore 1993).  By
contrast their predicted flux ratio was 1.1.

They investigated whether the difference in flux ratios might be due
to microlensing, assuming no smoothly distributed matter.  Their
{\it a posteriori} assessment was that the flux difference between the
observed optical flux ratio and the predicted ratio is ``unlikely, but
not ridiculously so.''

We carried out a second suite of simulations, M25 and S25 in Table 1,
using the WMS macro-parameters, but adding increasing proportions of
smoothly distributed matter, up to 97.5\%.

The magnification maps and histograms are similar in broad outline but
different in detail from the M10 and S10 models.  For the 0\% smooth
matter case, the caustic networks are very much denser.  The
histograms for the 0\% case are narrower than for models M10 and S10.
They are more nearly symmetric and are more similar to each other.

As the smooth matter percentage is increased both histograms widen.  The
low magnification tail for the macro-saddlepoint drops sharply at
$\mu_{\rm tot} \sim 1$, consistent with our toy model.  The saddlepoint
histogram bifurcates, with the bifurcation most prominent and the
distribution broadest for the 93\% smooth matter model.

\subsection{Lower Limits to the Magnification Histograms}
The magnification histograms in Figure 3 show sharp cutoffs at
faint magnification.  For saddlepoints this corresponds to
the minimum magnification in our toy model, equation (5),
which approaches unit magnification as $\kappa_{\rm tot} \approx \gamma$
approaches $\onehalf$.
For macro-minima the minimum magnification is given by
the magnification in the absence of the clumpy component,
\begin{equation}
\mu_{\rm min} = { 1
\over
[(1 - \kappa_c)^2 - \gamma^2]} \quad {\rm (minima)}
\end{equation}
(cf. Webster et al. 1991).  For minima, adding microlenses can only
increase the magnification above this value.

\subsection{Microlensing in the $\kappa^{\rm eff} - \gamma^{\rm eff}$ Plane}

The simulations presented in the preceding subsections trace locii in
the $\kappa^{\rm eff} - \gamma^{\rm eff}$ plane appropriate to images
with a constant observed magnification and   varying proportions of
smoothly distributed and microlensing matter.  It is worthwhile to consider the
results of previous investigations in this framework.  Watson and
Deguchi (1987, 1988) obtained analytic results along the $\gamma = 0$
axis, as did Seitz, Wambsganss and Schneider (1994).  Witt, Mao and
Schechter (1995) did simulations along the $\gamma = 3 \kappa - 1$ locus.
Wambsganss (1992) carried out         simulations for several
different values of $\gamma$ at $\kappa = 0.2$, $0.5$ and $1.5$.
Lewis and Irwin (1995) performed   simulations on an uneven $5 \times
5$ grid covering $0 < \kappa < 1$, $0 < \gamma < 1$.  They organized
their histograms so as to show their coverage of     the $\kappa -
\gamma$ plane.

Generalizing from the above results, it seems clear that microlensing
fluctuations go to zero along the locus of infinite magnification,
$\gamma = 1 - \kappa$, separating minima from saddlepoints.  Since
they also go to zero along the $\kappa = 0$ axis there must be a ridge of
maximum fluctuation amplitude (for minima) somewhere in between.
There would also appear to be a ridge for saddlepoints which has been
crossed by the WMS simulations and also by those in the preceding
section.  While the presently available coverage of the $\kappa -
\gamma$ plane is too sparse to accurately map out this locus,
it would seem to coincide, at least roughly, with the $\mu = 4$
hyperbola for minima and the $\mu = -3$ hyperbola for saddlepoints.
If so then the phenomenon of increasing fluctuations with the addition
of smoothly distributed matter is restricted to more highly magnified
images.

As is clear from Figures 3 and 4, the rms fluctuation in magnitude is
only one of several statistics needed to describe microlensing
magnification histograms in anything       like their full detail.
Their asymmetries vary considerably, and they often bifurcate.  We
would assign no special physical significance to the locus of maximum
rms fluctuation, taking it only to be a convenient measure of
variability.

\section{MICROLENSING AND MG0414+0534}

\subsection{The $A_2/A_1$ Flux Ratio}

Witt, Mao and Schechter (1995) present histograms of magnitude differences,
\begin{equation}
\Delta m_{A_1/A_2} \equiv -2.5 \log {A_1 \over A_2}
\end{equation}
for three values of the ratio of the Gaussian source radius, $r_s$ to
the Einstein ring radius $\xi_E$.  They estimate that the effect of
letting $r_s/\xi_E \rightarrow 0$ would broaden the histogram for
$r_s/\xi_E = 0.04$, their smallest value, by an additional 10\%.

For the sake of comparison with WMS, we carried out our M25 and S25
simulations using the same convergence and shear, and the same value
for the source radius, $r_s/\xi_E = 0.04$.                          
                                                               Figure
4 shows $\Delta m_{A_1/A_2}$ histograms computed for four of our
simulations (cases M25 and S25).  The first of these, with no smoothly
distributed matter, satisfactorily reproduces the WMS histogram, with small
differences consistent with differences in the simulation technique.
The second, with 87.5\% smooth matter, is considerably broader.
Moreover, it is asymmetric, with $A_2$ more likely to be fainter than
$A_1$.

The third, for 92.5\% smooth matter, shows the broadest distribution.
With still increasing fraction of smooth matter, the distributions
quickly get narrower again, as the fourth column for 97.5\% smooth
matter illustrates, though it does show a small plateau.

For the case of 0\% smooth matter, the probability that $A_2$ will be
more than 0.87 magnitudes fainter than $A_1$ is 0.068 (cf. WMS).  For
the 87.5\% smooth matter case, that probability has grown to 0.28, and
to 0.35 for the 93\% smooth matter case, an increase of more than a
factor of five.  We conclude that if the smoothly distributed matter
surface mass density along the line of sight were 87.5\% or 92.5\% of
the total, flux ratios as extreme as that seen in MG0414+0534 would
not be uncommon, as long as the saddlepoint image is the fainter of
the two images.

\subsection{Photometric Dark Matter Fraction}

On the assumption that the lensing galaxy in MG0414+0534 is like
nearby elliptical galaxies, a dark matter fraction can be derived as
follows.  Kochanek et al. (2000) measure a deVaucouleurs effective
radius for the lensing galaxy of 0\farcs78.  The Kormendy relation
(Kormendy 1977) presented by Bernardi et al. (2001) in their Figure 49
can be used to obtain an expected zero redshift surface brightness.
To take advantage of the mass to light ratios presented by Kauffmann et
al. (2002) we use the $g'$ surface brightness, for which the M/L for a
present day elliptical is roughly 4 if $H_0$ = 70 km/s/Mpc.  The
stellar surface mass density is then 386 $M_\sun$/pc$^{2}$ at the
observed effective radius of the lens, assuming no evolution
in the mass profiles of elliptical galaxies.

For an Einstein ring radius of $1\farcs15$ (Trotter et al. 2000) and a
deVaucoulers profile, the stellar surface density is smaller on the
ring by a factor of 0.46, giving 177 $M_\sun$/pc$^{2}$.  For
a lens redshift $z_L = 0.96$ and a source redshift $z_S = 2.64$ we
find a critical surface density of $\Sigma_c = 2189 M_\sun$/pc$^{2}$
for $(\Omega_m, \Lambda) = (0.3,0.7)$.  For an isothermal sphere the
surface density on the Einstein ring is one half the critical surface
density.  We therefore have ${\kappa_*/\kappa_{\rm tot} = 0.16}$ on
the Einstein ring, which is the approximate location of the $A_1$ and
$A_2$ images.  It is interesting to note that the ``best'' models of
Trotter et al. (2000) give magnifications for the $A_1$ and $A_2$ images
of order $\sim 15$, smaller by $3/5$ than the WMS values.  Had Witt et
al. adopted a model in which the quadrupole moment was due to
flattening of the galaxy rather than an external tide, they would have
found magnifications smaller by $\sim 1/2$, making our S10 and M10
models more appropriate.

\subsection{Smoothly Distributed Matter and Source Size}

Witt Mao and Schechter (1995) obtained a 95\% confidence upper limit
on the source size for MG0414+0534 of $10^{16}$ cm $ \times
(<M>/0.1M_\sun)^{1/2} $ for the case of no smoothly distributed
matter.  As increasing the smooth matter content increases the
fluctuation amplitude, that upper limit on the source size is relaxed.

Most simulations of microlensing use a finite source size,
parameterized by $r_s/\xi_E$.  This adds a third dimension to one's
model space.  A rough idea of the effect of source size can be had by
taking the source to be composed of two components -- one very
compact, so that point source simulations suffice, and one very
extended, so that the macro-magnification holds.  The resulting
magnification map is then just a luminosity weighted average of a
point source map and a uniform map.

The effect of increasing the extended fraction of a source is to
compress the magnification histograms of Figure \ref{fig-histo}
horizontally, preserving their shapes.  Thus while increasing the
extended fraction of the source would, for bright images, mimic
decreasing the smoothly distributed fraction of the lens, details of
the magnification histograms will be different and residuals from an
ensemble of identical systems would in principal permit further
tighter constraints on both the smoothly distributed mass fraction and
the extended source fraction.  The source and lens in MG0414+0534 are
surely moving with respect to each other.  If we wait long enough, we
will get a new, statistically independent set of flux ratios.
 
\subsection{Alternative Interpretations} 
 
Many gravitationally lensed systems have images with different
observed colors.  This could be a natural consequence of microlensing
under the assumption that the optical emission arises from an
accretion disk whose size is roughly that of the microlenses. Blue
flux coming from the inner part of the accretion disk would therefore
be more vulnerable to microlensing than red flux coming from further
out.  This would result in larger fluctuations (i.e. flux ratios) at
shorter wavelengths (Wambsganss and Paczy\'nski 1991).  In the case of
MG0414+0534, one might expect the blue flux to be more heavily
suppressed than the red flux.  In fact the $A_2/A_1$ flux ratio observed
at 2.05 \micron ~  {\it is} considerably closer to unity than that
observed at 0.8 \micron.

But dust in the intervening galaxy might also give rise to such an
effect.  Assuming that the color differences are due to dust in the
intervening galaxy, Falco et al. (1999) have used them to derive
corrected flux ratios, reducing all images to a common residual
extinction.  Adopting a standard extinction curve they derive an
extinction corrected $A_2/A_1 = 1.05$.  But Falco et al. call the
lensing galaxy an elliptical galaxy, and elsewhere the same authors
treat the lensing galaxy as elliptical (as most lenses are) including
it in their study of the fundamental plane of lensing galaxies
(Kochanek et al. 2000).  It seems unlikely to the present authors that
an E3 elliptical galaxy would have a single spot of dust with 1
magnitude of extinction at $1.5 r_e$, but as we are extrapolating from
our experience at $z=0$ to $z=0.96$, we cannot be absolutely certain
that we are seeing the effects of microlensing and not dust.

\section{THE SMOOTH MATTER FRACTION AS DETERMINED FROM ENSEMBLES}

One could in principal measure the smoothly distributed matter content
of a single lens by sampling the magnification histograms many times,
waiting for the source and lens to move with respect to each other,
given a macro-model with accurate values of the shear and convergence.  As
this could be inconvenient, an alternative would be to assemble an
ensemble of lenses and to look at the distribution of intensity ratio
residuals from best fitting models.  Several surveys for lenses are
underway that might produce such samples.  In particular the
Hamburg-ESO survey (Wisotzki et al. 1996) and the Sloan Digital Sky
Survey (Schneider et al. 2002) will ultimately yield large number of
lenses.

\subsection{Point Sources}
The treatment of an ensemble of lenses is most straightforward in the
limit where $r_s/\xi_E$ is small.  For every image in every quadruple
in the sample, one would obtain a model (ideally not using the flux
ratios as constraints) and compute magnification histograms for each
of the four images for a range of values of $\kappa_c/\kappa_{tot}$.
One would then construct a likelihood function from the product of the
probabilities obtained from the histogram.  There would be an
additional free parameter for each system -- a number which converts
observed flux into magnification.  Maximizing the likelihood would
give a best value for the smoothly distributed mass fraction,
$\kappa_c/\kappa_{tot}$.  

The number of systems needed to establish the presence of smoothly
distributed matter need not be large.  From the histograms in Figure 3
we see that a single system with a saddlepoint 2 magnitudes fainter
than predicted would rule out the 0\% smooth case at the $2 \sigma$
level.

A variant of this approach would be to assign a range of mass-to-light
ratios to the observed surface brightness profile for each lensing
galaxy, attributing the shortfall in surface density to smoothly
distributed matter.

\subsection{Finite Sources}

There are both theoretical and observational reasons to think that
$r_s/\xi_E$ is not vanishingly small, at least for some systems.  The
one-size-fits-all approach would be to assign a single value of
$r_s/\xi_E$ to every system (or a fixed {\em physical} size $r_s$,
scaled properly by the lens-dependent $\xi_E$), and then maximize the
likelihood function with this additional variable.  One might use a
Gaussian surface brightness profile for the sources or, for the sake
of computational speed, treat the quasar as a point source embedded in
a very extended source.  More realistically, but at the cost of
considerable additional complexity, one could allow each system to
have a different (e.g., luminosity-dependent) source size and profile.

\subsection{Double Image Systems}

We have emphasized quadruple systems rather than doubles because the
effects of smoothly distributed matter would seem to be more
important at high magnifications.  The results of the previous
sections would indicate that even doubles that are not highly magnified
could still suffer substantial microlensing.

However, doubles suffer from another difficulty -- a dearth of model
constraints.  One needs 5 constraints to obtain the simplest SIS with
external shear model, and doubles have only four positional
constraints -- the two image positions relative to the lens.  The flux
ratio is often taken as the fifth constraint, but using it as a
constraint will give a perfect fit to the model.  Doubles are not
beyond redemption, however; one might use radio flux ratios, emission
line ratios (Wisotzki et al. 1993), or mid-IR flux ratios (Agol et
al. 2000) to constrain the model on the hypothesis that the emission
regions are large compared to the Einstein rings of the
microlenses\footnote{Such observations might also be used as
additional constraints in models for quadruple systems.}.

\section{FURTHER CONSEQUENCES}

\subsection{Millilensing and Mini-halos}

The present work is fundamentally similar to (though on the surface
quite different from) that of Metcalf and Madau (2001), Chiba (2002)
and Dalal and Kochanek (2002) who argue that the brightness ratio
anomalies observed in quadruple systems are due to the {\it presence}
of dark matter mini-halos.

The differences are obvious.  They substitute increasing amounts of
clumpy dark matter for smoothly distributed luminous matter.  We
substitute increasing amounts of smoothly distributed ``dark'' matter for
clumpy luminous matter.  But the phenomenon we seek to explain is the
same.

We find, as they do, that a small micro- (or milli-) lensing
fraction produces surprisingly large flux ratio anomalies.
In our case we find this to be especially true for saddlepoints.

But the situation for extended micro- or millilenses is somewhat
different than for point mass perturbers.  If we follow the same line
of reasoning as in our toy model, but taking the perturbers to be
singular isothermal spheres directly along the line of sight to the
macro-images, we find that both the micro-minima and
micro-saddlepoints have twice the magnification that the corresponding
point mass perturber would have had.  In the limit as $\gamma \approx
\kappa_{tot} \rightarrow \onehalf$, the SIS perturber doubles the
magnification of a macro-minimum and reduces the macro-saddlepoint but
still leaves it a factor of two brighter than the unmagnified source.
We therefore expect extended perturbers to produce larger fluctuations
for macro-minima and smaller fluctuations for macro-saddlepoints than
would point mass perturbers.  The anomalies ought to be more uniformly
distributed among the multiple images, and not so heavily weighted
toward the saddlepoints.  Perhaps this is why Dalal and Kochanek
(2002) made no distinction among macro-minima and macro-saddlepoints
in their discussion (though they appear to have treated these
correctly in their simulations).  Metcalf and Madau (2001) do note an
asymmetry between the two, and their figures show it to be in the same
sense, if not as exaggerated, as those we find for point perturbers.

Dalal and Kochanek (2002) argue that the anomalous flux ratios
observed at {\it radio} wavelengths are due to millilensing by
subhalos comprised entirely of dark matter.  This rests on the fact that radio
sources are larger than the Einstein rings of individual stars.
Metcalf and Zhao (2001), building on the work of Metcalf \& Madau
(2001), argue for a similar effect including optical as well as radio
fluctuations in their discussion.

If radio emission regions in quasars were small compared to the
Einstein rings of the stars in the lensing galaxies, the results of
the previous sections would undermine the conclusions regarding
subhalos, since some fraction of the observed fluctuations would be
due to micro- rather than millilensing.

\subsection{Millilensing vs. Microlensing}

In fact, there is a way to distinguish  microlensing effects
that we propose here as an explanation for the discrepant 
intensity ratios in close double images, from millilensing,
as suggested by 
Mao \& Schneider (1998),
Metcalf \& Madau (2001), Metcalf \& Zhao (2001), or
Dalal \& Kochanek (2002).
Since quasar, lensing galaxy  and observer move relative to
each other, the brightness of one particular quasar image 
will change with time due to the fact that the focussing matter
in front of it changes position. The  time scale of
such fluctuations is of order a few years up to a decade
for microlensing by solar-mass stars. Since this timescale
is proportional to the Einstein radius of the lenses, 
it increases with the square root of the mass of the lensing
objects. 
Typical masses of substructure clumps range from
about $10^2 M_\odot$  to $10^{6-9} M_\odot$   
(Metcalf \& Madau 2001, Dalal \& Kochanek 2002).
This  means millilensing effects  of such substructure clumps
should not modify the intensity ratios of affected 
multiple-quasars over the professional lifetime of an astronomer,
whereas microlensing should produce variable intensity ratios.

\subsection{Modelling Lenses}

The magnification histograms presented in Figures \ref{fig-histo} and 
\ref{fig-A1-A2} show that
flux ratios are reliable constraints only in the absence of micro- and
millilensing.  If either is suspected, one will want to take the
expected fluctuations into account.  Since the histograms can be
quite broad and heavily skewed, proper treatment demands a
full-blown maximum likelihood analysis rather than the assumption of
Gaussian residuals.  Fitting for fluxes should work better than
fitting for magnitudes (i.e., log flux), since the mean magnitude
residuals can be quite different from zero.

\subsection{Non-isothermality: Is Smoothly Distributed Matter Necessary?}

The singular isothermal sphere model discussed here is the ``industry
standard'' for microlensing.  External evidence from the dynamics of
nearby elliptical galaxies (Romanowsky and Kochanek 1999) and internal
evidence from lensed systems with multiple and extended sources
(Kochanek 1995; Bernstein and Fischer 1999; Cohn et al. 2001; Rusin et
al. 2002) consistently give potentials with logarithmic slopes very
nearly that of an isothermal sphere.

It should be noted, however, that we can also obtain larger
fluctuation amplitudes for the bright pairs of images without smoothly
distributed matter, by going to models with stars alone that have
$\kappa_*$ and $\gamma$ equal to our effective convergences and
shears.  Witt, Mao and Schechter (1995) obtained macro-models for a
more centrally concentrated potential midway between a singular
isothermal sphere and a point mass (parameterized by setting
their $\beta = 1/2$).  This comes close to what one might expect for a
deVaucouleurs $r^{1/4}$ profile at constant mass to light ratio.

The magnifications are then smaller by roughly a factor of two than
for their isothermal models.  For MG0414+0534, the $\beta = 1/2$ model
gives convergences and shears for the bright pair of images of
(0.229,0.710) and (0.239,0.814), respectively, for the $A_1$ minimum and
$A_2$ sadddlepoint.  The isothermal model, with $\kappa_*/\kappa_{\rm tot} =
0.333$, would give an effective convergence and shear of (0.230,0.712)
and (0.239, 0.813), making the two fluctuation histograms virtually
indistinguishable.

If a very long sequence of observations ruled out $\kappa_*/\kappa_{\rm
tot} = 0.333$ for the isothermal model, it would also rule out the
$\beta = 1/2$ model.  At least in principle, then, fluctuation
observations might rule out constant mass to light ratios.

Taken alone, the qualitative aspects of anomalous flux ratios, and in
particular suppressed saddlepoints, cannot establish the presence of a
dark component.  But to the extent that other techniques do demonstrate
the presence of a dark component, anomalies can establish that this dark
component must be smooth.

\section{SUMMARY AND CONCLUSIONS}

We have shown that the substitution of a smoothly distributed matter
component for the clumpy stellar surface density {\it increases}
rather than decreases the amplitude of microlensing brightness
fluctuations for close pairs of images in quadruple gravitational
lenses.  We have also shown that in the presence of a smoothly
distributed component, the magnification probabily distributions
for saddlepoints and minima of the Fermat arrival time surface are
quite different, making it likely that saddlepoints undergo
substantial demagnification.
Inclusion of a smoothly distributed matter component in models for
MG0414+0534 makes the large optical flux ratio of the bright pair of
images considerably more likely under the microlensing hypothesis.

In extreme cases a single flux ratio anomaly might serve to establish
the relative proportions of smooth and microlensing mass.  More
generally the relative contribution of smoothly distributed matter may
be established either by observing an ensemble of lensed systems or by
carrying out a (long) time series for a single system.



\acknowledgments

We thank our colleagues who have graciously shared news of their
discoveries in advance of publication.  We thank the referee
for a careful reading and helpful suggestions.  We gratefully acknowledge
generous support from the Institute for Advanced Study and Princeton
University Observatory.  PLS is grateful to the John Simon Guggenheim
Foundation for the award of a Fellowship.

\clearpage


\begin{figure}
\begin{minipage}{165mm}
\plotone{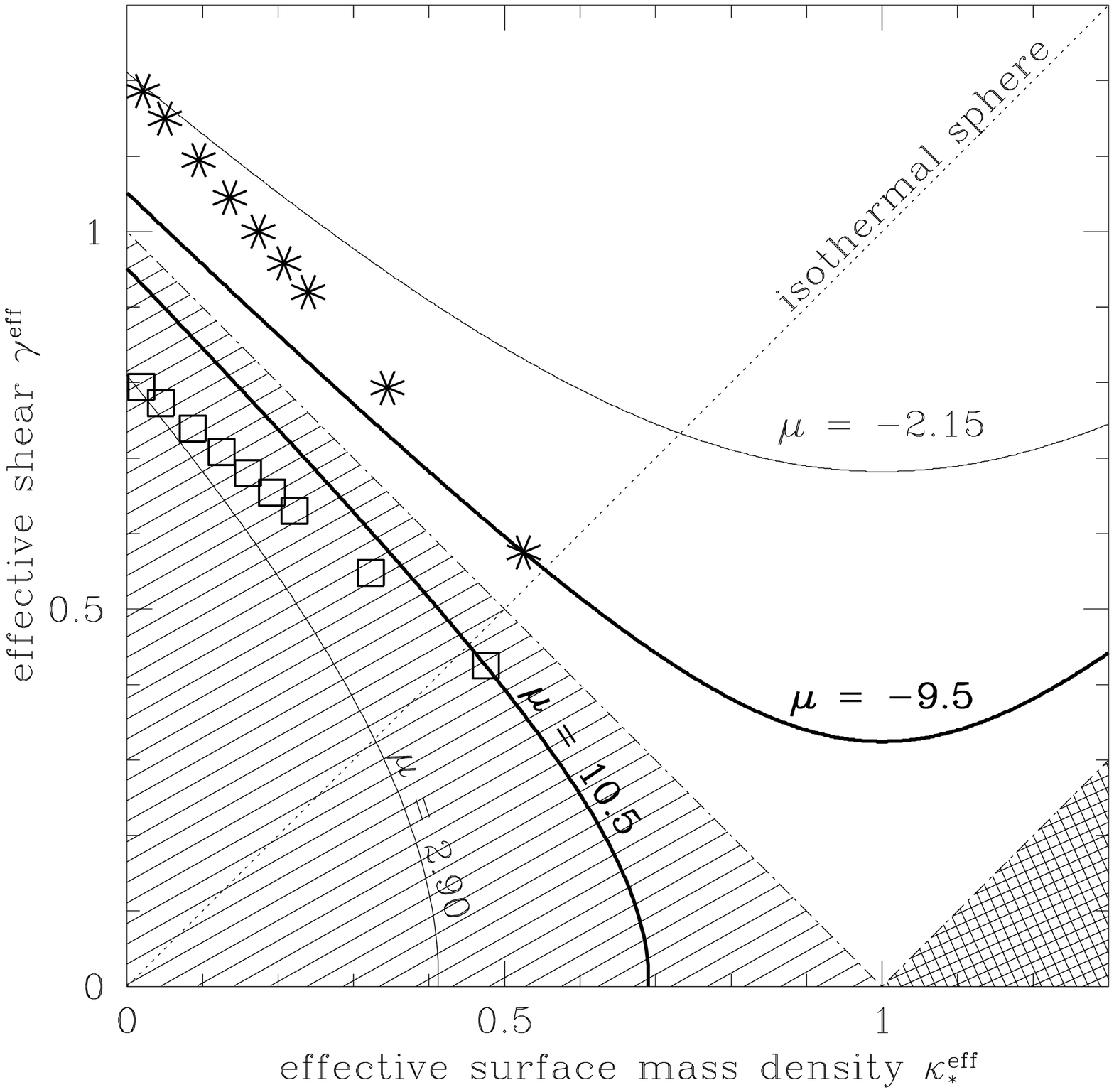}
\caption{$\kappa^{\rm eff}$-$\gamma^{\rm eff}$-plane.  Images
corresponding to minima in the arrival time surface form in the
hatched region; maxima form in the cross-hatched region; saddlepoint
images
form in the remaining triangular region.  Lines of constant
magnification are hyperbolae.  
The symbols indicate the models we investigated (M10 -- squares;
S10 -- stars). The symbols approach the
vertical axis as increasing amounts of smoothly distributed matter are substituted for
microlensing matter, keeping $\kappa_{tot}$ constant.
\label{fig-gam-kap}}
\end{minipage}
\end{figure}

\clearpage

\begin{figure}
\begin{minipage}{150mm}
\vspace{-30mm}
\plotone{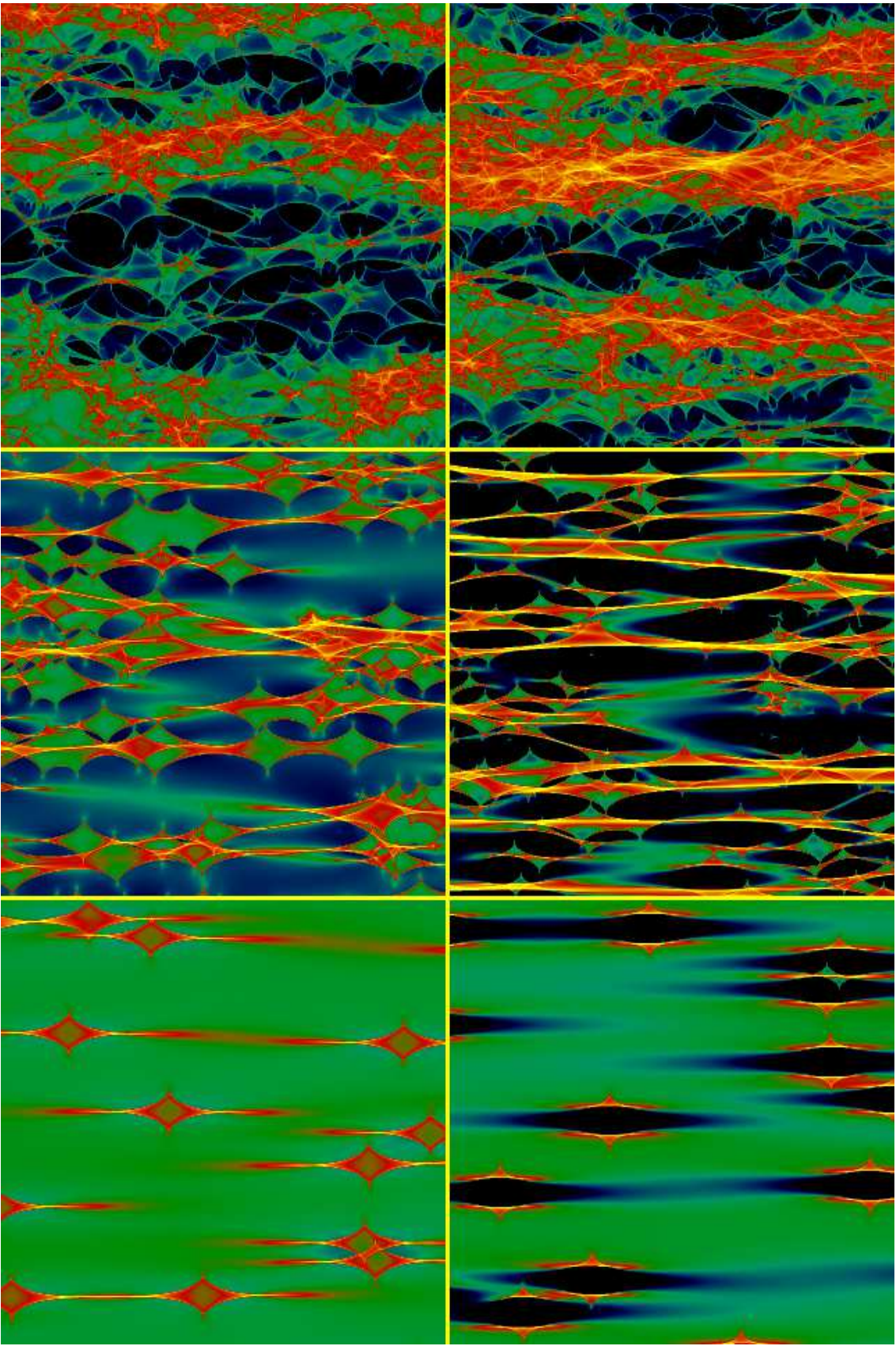}
\caption{Two-dimensional microlensing magnification distribution in
the quasar plane, for a minimum (left, models M10) and 
a saddlepoint image (right, models S10).  
The color scale ranges from dark blue (large
demagnification) - light blue - green - red - yellow (large
magnification).  The total convergence $\kappa_{tot}$ remains constant
for each column, whereas the fraction of smoothly distributed matter
increases downward from 0\% at the top through 85\% in the middle to
98\% at the bottom (corresponding to rows 1, 3 and 5 of 
Figure \ref{fig-histo}).
\label{fig-color}}
\end{minipage}
\end{figure}

\clearpage

\begin{figure}
\begin{minipage}{165mm}
\plotone{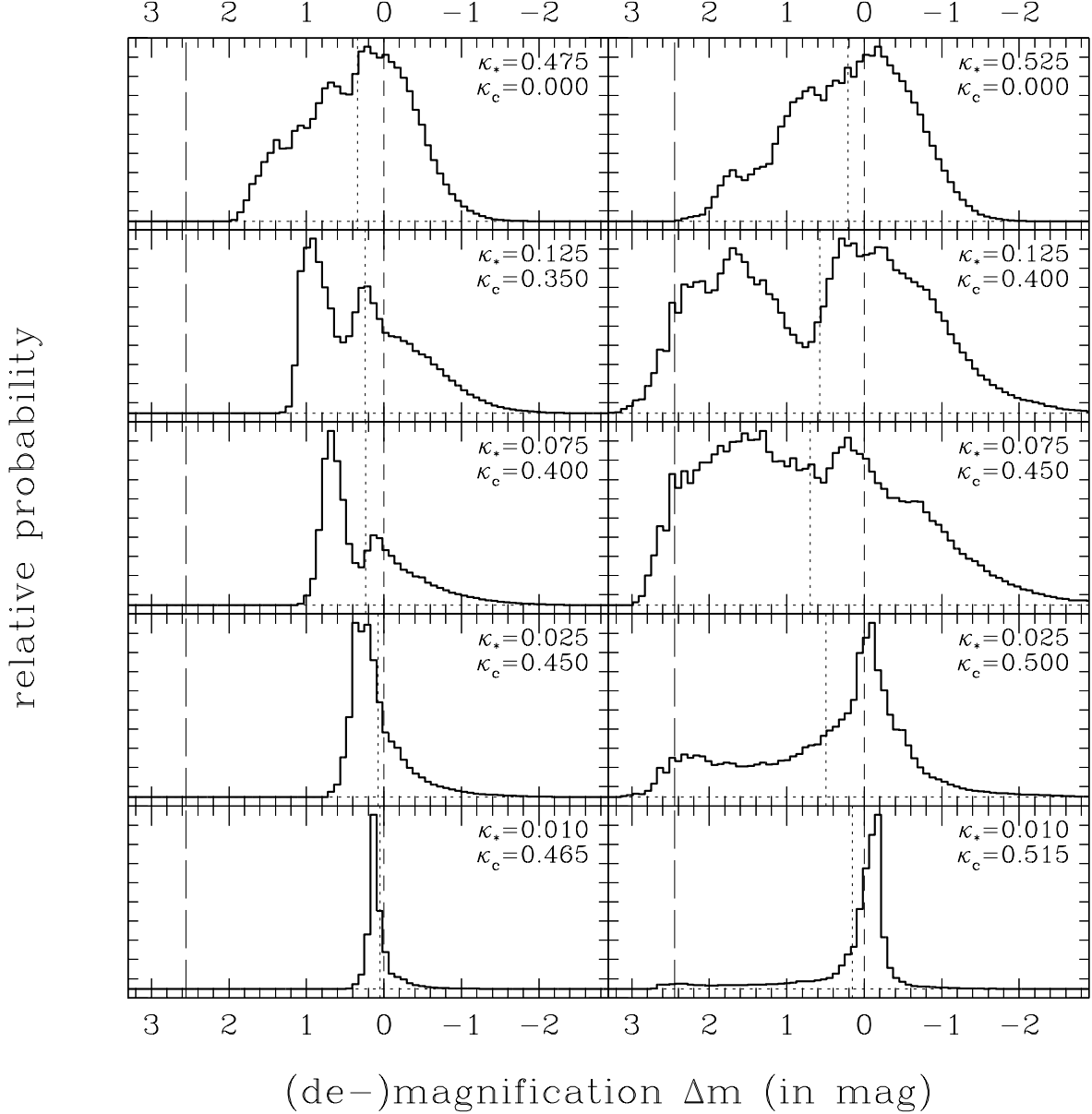}
\caption{Magnification probability distribution for a minimum (left,
models M10, $\gamma = 0.425$)
and a saddlepoint image (right, models S10, $\gamma = 0.575$).  
The total convergence $\kappa_{tot}$
remains constant for each column.  The smoothly distributed  matter increases from top
to bottom, with fractional contributions of 0\%, 75\%, 85\%, 95\% and
98\%, respectively.
The three vertical lines indicate the following:
short-dashed: $\Delta m = 0$ mag (theoretically expected  macro-magnification);
dotted: $<\Delta m>$ (average magnification in magnitudes);
long-dashed: $\mu_{\rm abs} = 1.0$ (absolute magnification unity, i.e.
unlensed case).
\label{fig-histo}}
\end{minipage}
\end{figure}

\clearpage

\begin{figure}
\begin{minipage}{165mm}
\vspace{-30mm}
\plotone{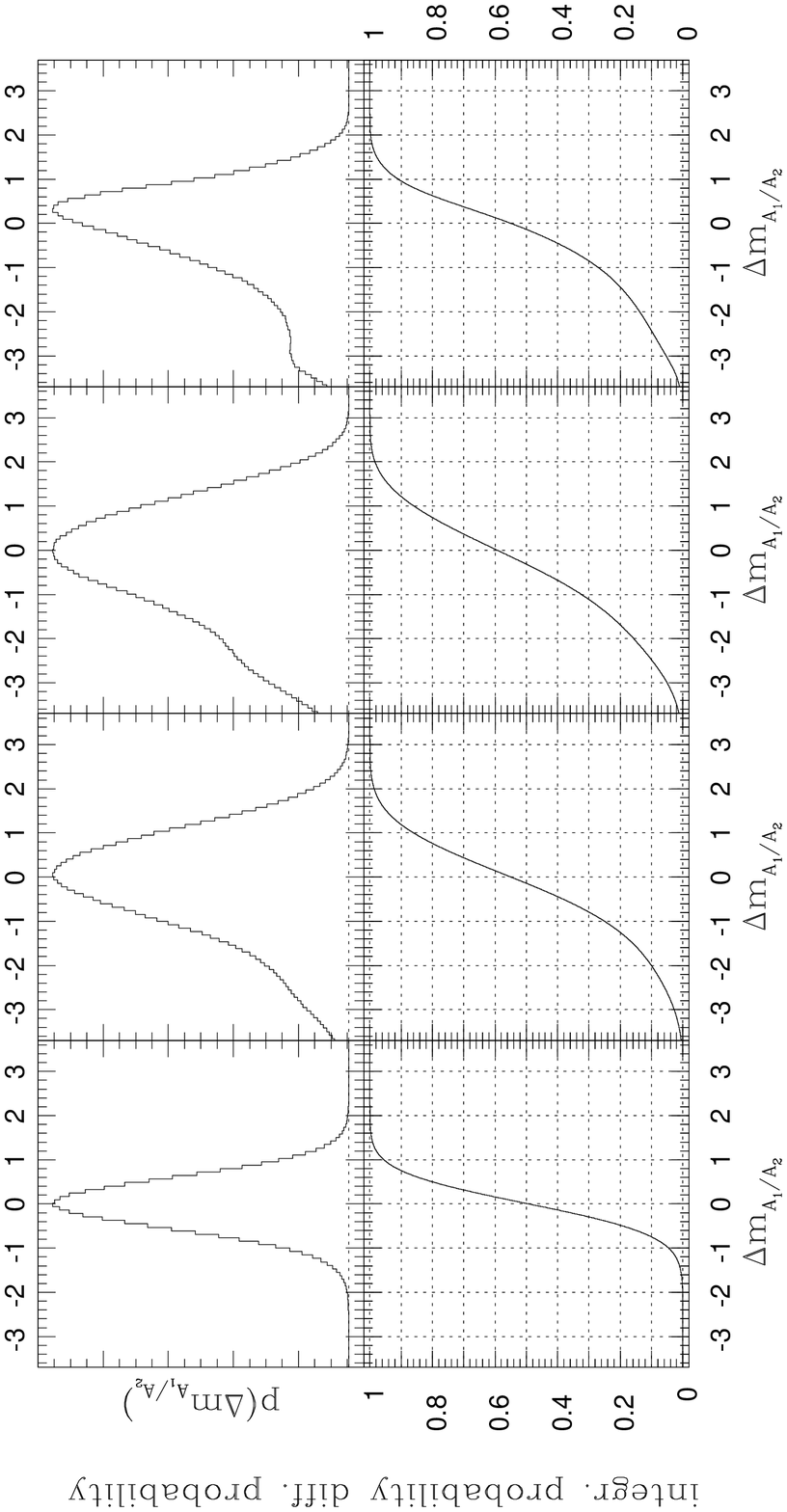}
\caption{Probability distributions
for the magnitude {\em difference} between images A$_1$ and A$_2$ of
MG0414+0534, with parameters as chosen by WMS:
$\kappa_{\rm A_1, tot} = 0.472$, $\gamma_{\rm A_1} = 0.488$
(our model M25), and 
$\kappa_{\rm A_2, tot} = 0.485$, $\gamma_{\rm A_2} = 0.550$
(our model S25).
Column 1 illustrates the case of {\em all matter}
in clumpy form (same as treated in Figure 4 of WMS); the 
panel  at the top shows the 
differential probability, the bottom panel presents the
corresponding integrated probability.
Columns 2, 3  and 4 show the corresponding
probability distributions for 88.5\% smooth matter,
  93\% smooth matter,
  and 
  97.5\% smooth matter, respectively.
\label{fig-A1-A2}}
\end{minipage}
\end{figure}









\begin{thebibliography}{}
\bibitem[aaa]{bbb} Agol, E., Jones, 
B., \& Blaes, O.\ 2000, ApJ, 545, 657 

\bibitem[aaa]{bbb} 
Bernardi, M. et al.\ 2001, preprint (astro-ph/0110344)

\bibitem[Bernstein \& Fischer(1999)]{1999AJ....118...14B} Bernstein, G.~\& 
Fischer, P.\ 1999, \aj, 118, 14 

\bibitem[aaa]{bbb}
Blandford, R.~D. \& Narayan, R.\ 1986, ApJ 310, 568


\bibitem[aaa]{bbb} Chang, K.~\& Refsdal, S.\ 1979, Nature, 282, 561 

\bibitem[aaa]{bbb} Chang, K.~\& Refsdal, S.\ 1984, A\&A, 132, 168 

\bibitem[Chiba(2002)]{2002ApJ...565...17C} Chiba, M.\ 2002, \apj, 565, 17 

\bibitem[Cohn, Kochanek, McLeod, \& Keeton(2001)]{2001ApJ...554.1216C} 
Cohn, J.~D., Kochanek, C.~S., McLeod, B.~A., \& Keeton, C.~R.\ 2001, \apj, 
554, 1216 

\bibitem[Dalal \& Kochanek(2002)]{2002ApJ...572...25D} Dalal, N.~\& 
Kochanek, C.~S.\ 2002, \apj, 572, 25 

\bibitem[aaa]{bbb}
Deguchi, S. \& Watson, W.~D.\ 1987, PRL 59, 2814

\bibitem[aaa]{bbb}
Deguchi, S. \& Watson, W.~D.\ 1988, ApJ 335, 67

\bibitem[Falco et al.(1999)]{1999ApJ...523..617F} Falco, E.~E.~et al.\ 
1999, \apj, 523, 617 

\bibitem[aaa]{bbb}
Gaudi, B.~S. \& Petters, A.~O.\ 2002, ApJ 574, 970

\bibitem[aaa]{bbb}
Hewitt, J.~N., Turner, E.~L., Lawrence, C.~R., Schneider, D.~P. 
	\& Brody, J.~P.\ 1992, AJ 104, 968

\bibitem[aaa]{bbb}
Kayser, R., Refsdal, S. \& Stabell, R.\ 1986, A\&A 166, 36

\bibitem[aaa]{bbb}
Katz, C.~A. \& Hewitt, J.~N.\ 1993, ApJ 409, L9

\bibitem[aaa]{bbb}
Kauffmann, G. et al.\ 2002, preprint (astro-ph/0204055)

\bibitem[aaa]{bbb}
Kochanek, C.~S.\ 2002, preprint (astro-ph/0204043)

\bibitem[Kochanek(1995)]{1995ApJ...445..559K} Kochanek, C.~S.\ 1995, \apj, 
445, 559 

\bibitem[Kochanek et al.(2000)]{2000ApJ...543..131K} Kochanek, C.~S.~et 
al.\ 2000, \apj, 543, 131 

\bibitem[Kormendy(1977)]{1977ApJ...218..333K} Kormendy, J.\ 1977, \apj, 
218, 333 

\bibitem[aaa]{bbb} 
Lawrence, C.~R., Elston, R., Januzzi, B.~T., \& Turner, E.~L.\ 1995, ApJ, 
110, 2570. 

\bibitem[aaa]{bbb}
Lewis, G.~F. \& Irwin, M.~J.\ 1995, MNRAS 276, 103

\bibitem[aaa]{bbb}
Lewis, G.~F. \& Irwin, M.~J.\ 1996, MNRAS 283, 225


\bibitem[aaa]{bbb}
Lewis, G.~F., Miralda-Escud{\'e}, J., Richardson, D.~C. 
	\& Wambsganss, J.\ 1993, MNRAS 261, 647 

\bibitem[aaa]{bbb}
Mao, S. \& Paczy\'nski, B.\ 1991,  ApJ 374, L37

\bibitem[aaa]{bbb}
Mao, S. \& Schneider, P.\ 1998, MNRAS 295, 587

\bibitem[aaa]{bbb}
Metcalf, R.~B. \& Madau, P.\ 2001, ApJ 563, 9 

\bibitem[aaa]{bbb}
Metcalf, R.~B. \& Zhao, H.\ 2001, ApJL 456, L5

\bibitem[aaa]{bbb}
Paczy{\'n}ski, B.\ 1986, ApJ 301, 503


\bibitem[aaa]{bbb}
Rauch, K.P., Mao, S., Wambsganss, J. \& Paczy{\'n}ski, B.\ 1992, 
	ApJ 386, 30 

\bibitem[aaa]{bbb}
Refsdal, S. \& Surdej, J.\ 1994, Rep. on Progr.  Phys. 56, 117

\bibitem[aaa]{bbb}
Reimers, D., Hagen, H.~J., Baade, R.,  Lopez, S. \& Tytler, T.\ 
	2002, A\&A, 382, L26

\bibitem[Romanowsky \& Kochanek(1999)]{1999ApJ...516...18R} Romanowsky, 
A.~J.~\& Kochanek, C.~S.\ 1999, \apj, 516, 18 

\bibitem[Rusin et al.(2002)]{2002MNRAS.330..205R} Rusin, D., Norbury, M., 
Biggs, A.~D., Marlow, D.~R., Jackson, N.~J., Browne, I.~W.~A., Wilkinson, 
P.~N., \& Myers, S.~T.\ 2002, \mnras, 330, 205 

\bibitem[Saha(2000)]{2000AJ....120.1654S} Saha, P.\ 2000, \aj, 120, 1654 

\bibitem[aaa]{bbb}
Schechter, P.~L. \& Moore, C.~B.\ 1993, AJ 105, 1

\bibitem[aaa]{bbb} Schneider, D.~P. et al.\ 2002, AJ, 123, 567 

\bibitem[aaa]{bbb}
Schneider, P. \& Weiss, A.\ 1987, A\&A 171, 49

\bibitem[aaa]{bbb}
Seitz, C. \& Schneider, P.\ 1994, A\&A 288, 1

\bibitem[aaa]{bbb}
Seitz, C., Wambsganss, J. \& Schneider, P.\ 1994, A\&A 288, 19

\bibitem[aaa]{bbb}
Trotter, C.~S., Winn, J.~N. \& Hewitt, J.~N.\ 2000, ApJ 535, 671


\bibitem[aaa]{bbb}
Wambsganss, J.\ 1990, PhD Thesis, Report MPA550

\bibitem[aaa]{bbb}
Wambsganss, J. 1992, ApJ 386, 19

\bibitem[aaa]{bbb}
Wambsganss, J.\ 1997, MNRAS 284, 172

\bibitem[Wambsganss \& Paczynski(1991)]{1991AJ....102..864W} Wambsganss, 
J.~\& Paczy\'nski, B.\ 1991, \aj, 102, 864 

\bibitem[aaa]{bbb}
Wambsganss, J., Witt, H.~J. \& Schneider, P.\ 1992, A\&A 258, 591

\bibitem[aaa]{bbb}
Wambsganss, J., Paczy\'nski B. \& Schneider, P.\ 1990, ApJ 358, L33

\bibitem[Webster, Ferguson, Corrigan, \& Irwin(1991)]{1991AJ....102.1939W} 
Webster, R.~L., Ferguson, A.~M.~N., Corrigan, R.~T., \& Irwin, M.~J.\ 1991, 
\aj, 102, 1939 

\bibitem[aaa]{bbb}
Wisotzki, L., K{\"o}hler, T., Kayser, R. \& Reimers, D.\ 1993, A\&A 278, L15

\bibitem[aaa]{bbb}
Wisotzki, L., K{\"o}hler, T., Groote, D., \& Reimers, D.\ 1996, A\&A,
115, 227

\bibitem[aaa]{bbb}
Witt, H.~J.\ 1993, ApJ 403, 530

\bibitem[aaa]{bbb}
Witt, H.~J., Mao, S. \& Schechter, P.~L.\ 1995, ApJ 443, 18 (WMS)
\bibitem[aaa]{bbb} Wucknitz, O.\ 2002, MNRAS 332, 951

%
%
%
%

\end{thebibliography}
\end{document}